\documentclass[12pt]{article}

\usepackage[title]{appendix}

\usepackage{aas_macros}
\usepackage{geometry}                
\usepackage{fullpage}  
\usepackage{url}

\usepackage[scaled=0.875]{helvet} 

\usepackage[OT1,T1]{fontenc}

\usepackage[colorlinks=true, linkcolor=blue, backref=page]{hyperref}

\usepackage{ifthen} 
\newboolean{enable-backrefs} 
\setboolean{enable-backrefs}{true} 

\newcommand{\backrefnotcitedstring}{\relax} 
	\newcommand{\backrefcitedsinglestring}[1]{(Cited on page~#1.)}
	\newcommand{\backrefcitedmultistring}[1]{(Cited on pages~#1.)}
	\ifthenelse{\boolean{enable-backrefs}} 
{
\PassOptionsToPackage{hyperpageref}{backref}
\usepackage{backref} 
\renewcommand*{\backref}[1]{}  
\renewcommand*{\backrefalt}[4]{
\ifcase #1 
\backrefnotcitedstring
\or
\backrefcitedsinglestring{#2}
\else
\backrefcitedmultistring{#2}
\fi}
}{\relax}

\usepackage[authoryear]{natbib}

\usepackage{graphicx}

\newcommand{\pcm}{\,cm$^{-3}$}

\title{Cosmic-ray pressure driven magnetic field amplification: dimensional, radiative and field orientation effects.}
\author{T. P. Downes $^{1,2,3}$ and L. O'C. Drury$^{2}$ \\
		\url{mailto:turlough.downes@dcu.ie} \\
		\url{mailto:ld@cp.dias.ie}\\[2ex]
$^{1}$School of Mathematical Sciences, Dublin City University, \\
	      Glasnevin, Dublin 9, Ireland\\[1ex]
$^{2}$School of Cosmic Physics, Dublin Institute for Advanced Studies, \\
31 Fitzwilliam Place, Dublin 2, Ireland \\[1ex]
$^{3}$National Centre for Plasma Science and Technology, \\
Dublin City University, Glasnevin, Dublin 9, Ireland}

\defcitealias{2012MNRAS.427.2308D}{Paper I}

\begin{document}
\maketitle
\sloppy

\begin{abstract}
Observations of non-thermal emission from several supernova remnants suggest
that magnetic fields close to the blastwave are much stronger than would be
naively expected from simple shock compression of the field permeating the
interstellar medium (ISM).  We investigate in some detail a simple model based on turbulence 
generation by cosmic-ray pressure gradients.  Previously this model was investigated using 2D
MHD simulations.

Motivated by the well-known qualitative differences between 2D and 3D turbulence, we further our 
investigations of this model using both 2D and 3D simulations to study the influence of the 
dimensionality of the simulations on the field amplification achieved.  Further, since the
model implies the formation of shocks which can, in principle, be efficiently cooled by
collisional cooling we include such cooling in our simulations to ascertain whether it could
increase the field amplification achieved.  Finally, we examine the influence of
different orientations of the magnetic field with respect to the normal of the blastwave.

We find that dimensionality has a slight influence on the overall amplification achieved, but a
significant impact on the morphology of the amplified field.  Collisional cooling has
surprisingly little impact, primarily due to the short time which any element of the ISM
resides in the precursor region for supernova blastwaves.  Even allowing for a wide range of
orientations of the magnetic field, we find that the magnetic field can be expected to be
amplified by, on average, {\em at least} an order of magnitude in the precursors of supernova
blastwaves.

\end{abstract}

\section{Introduction}
\label{sec:introduction}

The properties of the non-thermal emission observed in various supernova 
remnants suggest that the strength of magnetic fields in the vicinity of
the blastwave is significantly more than would be expected from
either the typical field strength in the interstellar medium (ISM), or the strength of that which would
be achieved by simple shock compression \citep{2003ApJ...584..758V, 2003A&A...412L..11B,
2004ApJ...602..257B, 2005ApJ...621..793B, 2005A&A...433..229V, 2006AdSpR..37.1902B,
2006A&A...453..387P,2012A&ARv..20...49V,Uchiyama:2007ly,2004A&A...416..595Y}.  The presence
of such amplified fields could fill in a missing piece in the theory of cosmic ray acceleration by 
SNR shocks and allow acceleration to the energies needed to explain the cosmic-ray `knee' 
particles; if the magnetic field strength is only a few $\mu$G as expected in the interstellar 
medium it is very hard to get the acceleration to reach these energies as pointed out by 
\cite{1983A&A...125..249L}.

We must then try to understand what physical process might give rise to such field
amplification.  An attractive idea is that in which magnetic fields might be amplified to
the required degree by cosmic-ray driven processes themselves, specifically those processes
occurring in strong shocks bounding young supernova remnants (SNRs). Indeed, in the work of
\cite{2004MNRAS.353..550B} it was pointed out that the existence of a strong
current-driven instability under conditions thought to be appropriate to
young remnants can amplify magnetic fields.  The Bell instability appears to be capable of
acting in the precursor regions of SNR shocks but unfortunately only acts on short
length-scales and, in the absence of some inverse cascade of energy, thus cannot create the
necessary relatively large length-scale field amplification required to explain the
observations.  Another candidate is the instability identified in \cite{Drury:1986uq} and further 
studied in \cite{Begelman:1994fj} ({\it cf} also \cite{Webb:1999vn} and \cite{Ryu:1993rt}).  This
has the advantage of operating on scales large compared to the gyro-radius of the driving particles 
and in relying only on rather simple and robust physics.

In \citet[hereafter Paper I]{2012MNRAS.427.2308D}, we presented a simple 
model in which the cosmic ray pressure gradient drives turbulence in the 
precursor of the blastwave through inducing differential acceleration on the 
inhomogeneous ISM through which the blastwave and precursor are moving, similar to an idea
proposed by \citet{2009ApJ...707.1541B}.  This differential acceleration occurs by virtue of the 
fact that the ISM contains density variations resulting from
pre-existing turbulence. The cosmic ray pressure gradient exerts a body force on the ISM which 
is independent of density and the resulting differential acceleration induces 
strong shear in the flow which, through fluid instabilities, creates 
turbulence.  This turbulence stretches and twists the magnetic field 
permeating the ISM creating enhanced field strengths throughout the precursor 
and, indeed, in the medium behind the supernova blastwave.

A particularly attractive property of this model is that it acts on 
large length scales - lengths comparable to the diffusion lengths of the 
highest energy non-thermal particles accelerated at the blastwave.  Thus,
the amplification of the magnetic fields resulting from this process will
occur on roughly these length scales also and, therefore, have the potential to
explain the observations of narrow filaments of non-thermal emission in the
X-ray \citep{2012A&ARv..20...49V}.

In \citetalias{2012MNRAS.427.2308D} we presented 2D simulations in which we examined the action of 
the proposed process to ascertain whether or not it could plausibly produce
the required magnetic field strengths.  It was found that, indeed, 
significant field amplification can occur, up to at least a factor of 20 for
the conditions studied.  In this work we extend the exploration of this model
in several directions as follows:
\begin{itemize}
\item Turbulence in three dimensions (3D) is fundamentally different to 
two dimensional (2D) turbulence in that the energy cascade in 3D is from 
large length scales to small ones, while in 2D the reverse is the case. It is
important to ascertain whether or not this influences the likely relevance
of the process proposed in our toy model.
\item While radiative cooling through line emission is unimportant in the
blastwaves of young SNRs, the differential acceleration produced under the
proposed process will yield shock strengths which produce conditions under
which such radiative cooling may well be significant.  Such cooling will
change the nature of the density inhomogeneities, and therefore the 
differential acceleration, in the precursor and so might also effect the 
field amplification achieved.
\item The angle of the mean magnetic field to the normal of the blastwave
is an important factor in amplification.  One may expect that a perpendicular
shock will result in much stronger field amplification than a parallel shock. Note that we 
do not include here the effects of the angle of the field on the diffusion coefficient for the 
cosmic ray particles.  This is undoubtedly an important effect, but its influence is hard to 
gauge given the highly tangled nature of the magnetic field in the precursor region.
\end{itemize}

\citet{2013MNRAS.436..294B} have also investigated the model presented in
\citetalias{2012MNRAS.427.2308D}, mainly focusing on how the model might work for cosmic ray
precursors of intracluster shocks.  This latter work employed both 2D and 3D simulations to
investigate the amplification factors achieved for low Mach number ($M \approx 2$ -- 3) and high Mach
number ($M = 100$) shocks.  The results confirmed our original results.  Interestingly, this
work detected no major difference in the field amplification achieved in 2D and 3D simulations.
We return to this point in Sect.\ \ref{sec:results-dim}.

The layout of this paper is as follows: in section \ref{sec:model} we briefly describe the 
toy model introduced in \citetalias{2012MNRAS.427.2308D}; in section \ref{sec:method} we describe our 
approach to studying the model; while in section \ref{sec:results} we address each of the issues 
raised above in turn.  Finally we give our conclusions in section \ref{sec:conclusions}.

\section{Recapitulation of the toy model}
\label{sec:model}

The instability which is the basis for the proposed mechanism of amplifying the
magnetic field in the precursor of SNR shocks comes about because the cosmic ray
pressure gradient in this precursor exerts a ponderomotive force which is independent
of density on the ISM passing through the precursor.  Any 
pre-existing density fluctuations in the ISM will thus result in differential 
acceleration, inducing shear which then creates turbulence through various
instabilities such as the Kelvin-Helmholtz instability.  This turbulence then
amplifies the pre-existing magnetic field in the usual way through stretching and
folding the field lines.  Thus the whole system acts as a dynamo, converting the
energy of the cosmic rays into magnetic energy in the precursor region.

In order to attempt to model this process we make the rather drastic assumption that
the cosmic ray propagation is completely decoupled from the matter dynamics.  While
this assumption clearly omits a lot of important physics, it will nonetheless allow
us to model the system to give some insight into how such a process could influence
the strength of the magnetic field.  In fact, this assumption is likely to result in
an {\em underestimate} of the efficacy of the process in amplifying the magnetic
field as cosmic ray particles are likely to diffuse more effectively along channels
already evacuated by increased cosmic ray pressure, resulting in an enhanced
instability and therefore stronger field amplification.

We assume a linear cosmic ray pressure in the precursor and consider a system
consisting of a rectangular computational box extending in
the $x$-direction from $0$ to $L$ within which the cosmic ray pressure $P_{\rm C}$
rises linearly from zero at the inflow side to a
value of order the ram pressure of the inflowing plasma at the outflow side.  The shock 
position is thus taken to be at $x=L$ and
\begin{equation}
P_{\rm C}(x) =  \theta \rho_0 U_0^2 {x\over L},
\label{eqn:pc_def}
\end{equation}
where $0<\theta<1$ is a positive parameter less than unity.

As a result of this pressure gradient the flow is decelerated by a uniform body force
$-\theta\rho_0U_0^2/L$.  The flow is seeded with small-scale density fluctuations
with a log-normal distribution, which is what would be expected from isothermal
turbulence in the ISM. 

Presuming the incoming flow contains density irregularities of magnitude $\delta
\rho$ on a length scale $\lambda$ the bulk force, operating on a time scale of order
the advection time through the precursor, will generate velocity fluctuations of
magnitude
\begin{equation}
\delta u \approx {\delta \rho\over \rho_0} {1\over\rho_0} {\theta \rho_0 U_0^2\over
	L} {L\over U_0} \approx {\delta\rho\over\rho_0} \theta U_0
	\label{eqn:delta-u}
	\end{equation}
	on the same length scale $\lambda$.  If this is to drive turbulence we require the
	eddy turn-over time to be short compared to the outer-scale and thus
	\begin{equation}
{\lambda\over\delta u} \ll {L\over U_0} \Rightarrow \lambda \ll \theta
{\delta\rho\over\rho_0} L
\label{eqn:lambda-limit}
\end{equation}
Density fluctuations satisfying this not very restrictive condition should be capable
of inducing turbulence and thus magnetic field amplification.    The total amount of
kinetic energy available in the turbulence can be roughly estimated as

\begin{equation}
\label{eqn:e_f}
e_{\rm F} = {1\over 2}\rho_0 (\delta u)^2 \approx {1\over 2 \rho_0
}\left(\delta\rho\right)^2 \theta^2 U_0^2
\end{equation}
and thus the maximum amplified field should be below full equipartition by a
factor of order $\theta^2 (\delta\rho/\rho_0)^2$.  If nonlinear effects drive the
density fluctuations to saturation at $\delta\rho\approx \rho$ (as is probable) then
this process could be very efficient at converting flow energy into magnetic energy
if $\theta \approx 1$.

\section{Numerical method}
\label{sec:method}

As in \citetalias{2012MNRAS.427.2308D}, we use the {\it HYDRA} code
\citep{OSullivan:2006zr,OSullivan:2007mz}, configured for simulating ideal MHD flows
rather than multifluid MHD systems, to study this model.  The equations solved are
\begin{eqnarray}
\frac{\partial \rho}{\partial t} + \mathbf{\nabla} \cdot \left(\rho
  \mathbf{u}\right)  & = & 0 \label{mass} \\
\frac{\partial \rho \mathbf{u}}{\partial t} + \nabla\cdot\left( \rho \mathbf{u} 
  \mathbf{u} + P \mathbf{I}\right) & = & \mathbf{J}\times\mathbf{B}
	+ \mathbf{F}_{\rm cr}, \label{neutral_mom} \\
\frac{\partial e}{\partial t} + \mathbf{\nabla} \cdot \left[ \left(e +
   P\right)\mathbf{u}\right] & = &
	\mathbf{J}\cdot(\mathbf{u}\times\mathbf{B}) + \mathbf{F}_{\rm cr}\cdot\mathbf{u}
	\nonumber \\
		& & - n^2 \Lambda (T) \label{eqn:energy}\\
\frac{\partial \mathbf{B}}{\partial t} + \nabla\cdot(\mathbf{u}\mathbf{B}-
		\mathbf{B}\mathbf{u}) & = &0 , \label{B_eqn} \\
\nabla\cdot\mathbf{B} & = & 0 \label{divB}
\end{eqnarray}
where $\rho$ is the mass density, $n$ is the number density, $\mathbf{u}$ is the fluid velocity, $P$ is the 
thermal pressure, $\mathbf{B}$ is the magnetic field, $\mathbf{F}_{\rm cr}$ is the 
force due to the cosmic ray pressure gradient and $\mathbf{I}$ is the identity matrix.  
$\mathbf{F}_{\rm cr}$ is given by
\begin{eqnarray}
\mathbf{F}_{\rm cr} & = & - \mathbf{\nabla} P_{\rm C} \nonumber \\
                    & = & - \frac{\theta \rho_0 U_0^2}{L}\mathbf{\hat{\i}} 
\end{eqnarray}
\noindent (see equation \ref{eqn:pc_def}).

The final term on the right hand side of equation \ref{eqn:energy} is that for
radiative losses due to collisional cooling.  Implicit in using this functional form
for the cooling is that the ISM is dominated by atomic hydrogen which is fully
ionised.  There is some inaccuracy in this approximation.  However, we deal with this
as follows.  First, a fit was performed to the data in \citet{1993ApJS...88..253S} for
non-equilibrium cooling for a gas with solar abundances.  This fit was for a 6$^{\rm th}$ 
order polynomial.  This high order was found to be necessary to capture sufficiently
accurately (see below) one of the most important properties of the function: its slope as a
function of temperature.  The resulting cooling function was tested using the
overstable radiative shock test \citep[e.g.][]{1988ApJ...329..927G} and modified slightly in order to bring the
test results into agreement with the published data for this test.  Thus $\Lambda$ is
calculated as follows.  Set $\alpha = \log_{10} T$ and 
\begin{eqnarray}
g(T) = -0.651597 \alpha^6 + 20.2803 \alpha^5 -261.438 \alpha^4 + 1786.93
   \alpha^3 \nonumber \\
- 6831.53 \alpha^2 + 13856.8 \alpha - 11678.9.
\end{eqnarray}
\noindent Then 
\begin{equation}
\label{eqn:Lambda}
\Lambda (T) = 10^{g(T)}.
\end{equation}

These equations are advanced in time using a standard van Leer-type second order, 
finite volume, shock capturing scheme.  The magnetic field divergence is controlled 
using the method of \cite{Dedner:2002fr}.  The unusual form of the MHD equations used here
is due to {\it HYDRA} being a multifluid code, making this form of the equations 
more convenient.  This code has been extensively validated for both multifluid and ideal 
MHD set-ups \citep{OSullivan:2007mz,OSullivan:2006zr} and see also Appendix \ref{sect:appendix}.  

For the simulations presented in this work we take $\theta$ (defined in equation 
\ref{eqn:pc_def}) to be 0.6 which is observationally reasonable \citep[e.g.][and 
references therein]{2012A&ARv..20...49V}.  For the large Mach numbers associated with 
supernova blastwaves this will give us a very significant acceleration of the pre-shock 
flow.

\subsection{Set-up of the problem}
\label{sec:init-cond}

We proceed as in \citetalias{2012MNRAS.427.2308D}, formulating the problem in the rest frame of the blastwave
which is assumed to be planar.  For the purposes of this work we present our initial
conditions in physical, rather than dimensionless units.  The principle reason for
this is that we will be studying the effects of radiative (line) cooling on the flow
and hence it makes more sense to leave aside dimensionless parameters.  The speed of the 
blastwave is taken to be $2.9\times10^8$\,cm\,s$^{-1}$ and the mean temperature of the ISM 
is taken to be $10^4$\,K, assuming solar abundances.  This yields a Mach number for the blastwave of 290.  The position 
of the blastwave is at $x=L$ where we take $L = 5\times10^{17}$\,cm for this work.  The size
of the computational domain is $L\times \frac{L}{8}\times\frac{L}{8}$ (for 3D
simulations) and $L\times \frac{L}{8}$ for 2D simulations.  Thus the blastwave is
positioned at the boundary of our domain, and is not actually part of the
calculations at all, as is appropriate since we are interested only in the precursor
region.  The initial magnetic field in the ISM is taken to be uniform and of strength 
$3$\,$\mu$G while the mean density of the ISM is set at 
$\rho_0 = 2.3\times10^{-22}$\,g\,cm$^{-3}$, or 100\,cm$^{-3}$ unless otherwise stated.  The density 
distribution in the ISM is prescribed as in \citetalias{2012MNRAS.427.2308D}, having an RMS variation of 
0.2 and a log-normal distribution.  The pressure is taken to be uniform, yielding the mean temperature 
given above.

Using the resolution study presented in \citetalias{2012MNRAS.427.2308D} we know that, while this system
cannot be fully resolved unless simulations resolving the dissipation scale (either
viscous or resistive or both) are performed, a resolution of $2000\times250$ is
adequate for this set-up.  Hence all 2D simulations presented here have a resolution
of $2000\times250$ unless otherwise stated, while 3D simulations have a resolution of
$2000\times250\times250$.  The only case one might be concerned about is that where
radiative cooling is present.  Cooling introduces new length and time-scales and
these must be well resolved in order to be reasonably confident of the results of the
simulations.  Therefore we performed extensive validation simulations in 2D with
resolutions up to $4000\times500$ for the radiatively cooled simulations and found
that, indeed, simulations including this effect can be adequately resolved at
$2000\times250$.  Note, however, that if a higher ISM density is used this conclusion
will no longer be valid since the cooling length goes as the inverse of the square of
the density.

In order to study the influence of radiative cooling, the dimensionality of the
system and the influence of the angle of the field to the shock normal we performed a
suite of simulations as detailed in Table \ref{table:nom}.  Note that the field angle
is measured with respect to the normal of the blastwave.


\begin{table*}
\begin{minipage}{126mm}
\caption{Summary of the simulations presented in this work \label{table:nom}}
\begin{tabular}{lcccl} \hline
Simulation & Field angle & Dimension & Grid size & Notes \\ \hline
adb-2d-std-lr & $\frac{\pi}{2}$ & 2D & $1000\times125$ & Adiabatic \\
adb-2d-std & $\frac{\pi}{2}$ & 2D & $2000\times250$ & Adiabatic \\
adb-2d-std-hr & $\frac{\pi}{2}$ & 2D & $4000\times500$ & Adiabatic \\
adb-3d-std & $\frac{\pi}{2}$ & 3D & $2000\times250\times250$ & Adiabatic \\
cool-2d-std & $\frac{\pi}{2}$ & 2D & $2000\times250$ & Radiatively cooled \\
cool-2d-hden & $\frac{\pi}{2}$ & 2D & $2000\times250$ & Radiatively cooled, $\rho_0=10^3$\pcm \\
cool-3d-std & $\frac{\pi}{2}$ & 3D & $2000\times250\times250$ & Radiatively cooled \\
adb-2d-b00 & $0$ & 2D & $2000\times250$ & Adiabatic \\
adb-2d-b30 & $\frac{\pi}{6}$ & 2D & $2000\times250$ & Adiabatic \\
adb-2d-b60 & $\frac{2\pi}{6}$ & 2D & $2000\times250$ & Adiabatic \\
\end{tabular}
\end{minipage}
\end{table*}

\subsection{Boundary conditions}
\label{sec:boundary-conditions}

The boundary conditions are set as follows:
\begin{itemize}
\item The left-hand boundary, where pre-shock ISM flows into the computational domain, is a
forced boundary with velocity $2.9\times10^8$\,cm\,s$^{-1}$.  The density here is time
dependent and defined such that the random density distribution flowing onto the grid
matches that on the grid at $t=0$ and has a (spatial) periodicity of $L$.
\item The right-hand boundary, where pre-shock ISM flows off the grid, is set to gradient
zero boundary conditions.  This is appropriate as the ISM is still flowing supersonically at
this point and so any waves emitted from this boundary as a result of the imposition of the 
gradient zero boundary conditions will be carried off the grid.
\item All other boundaries are set to periodic.
\end{itemize}

\noindent With these boundary conditions we are effectively simulating an infinite domain in
the $y$ (and $z$) directions, with the restriction that wavelengths longer than $L/8$ can not
be represented.

\section{Results}
\label{sec:results}

\begin{figure*}
\begin{center}
\includegraphics[width=15cm]{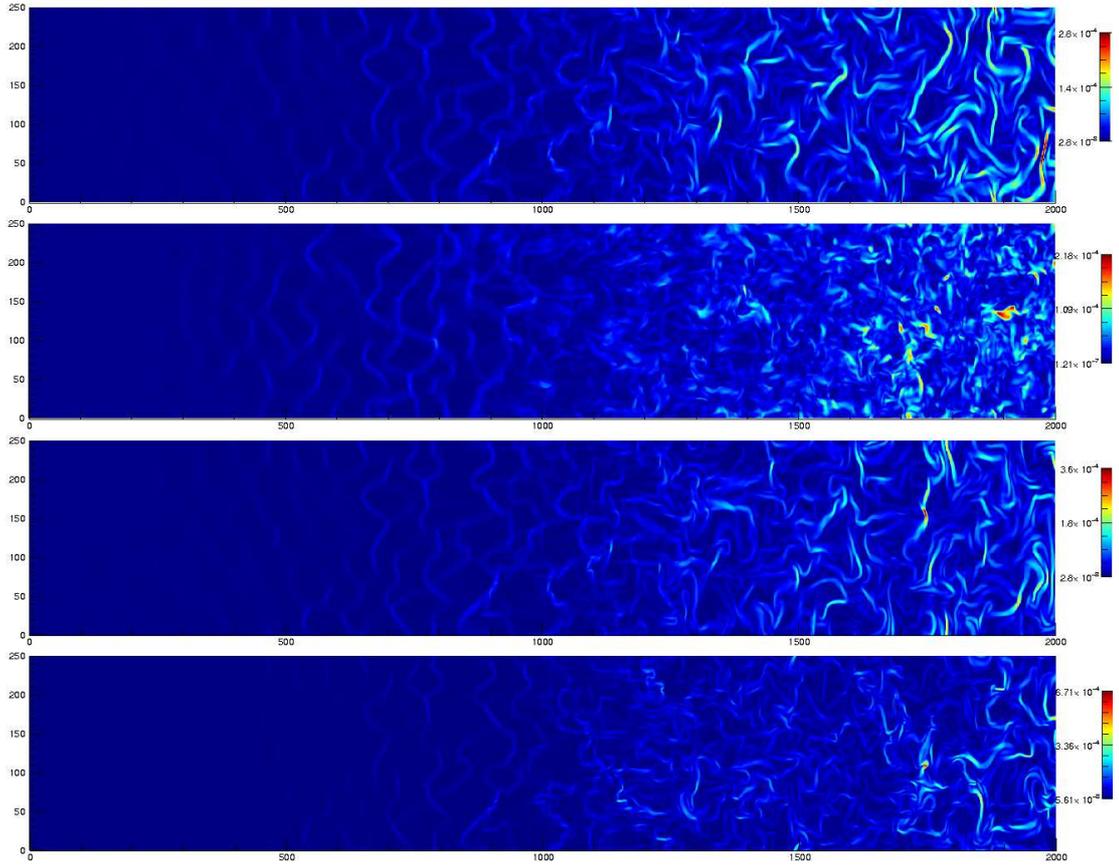}
\caption{ \label{fig:bmag-image} Plots of the magnetic field strength for adb-2d-std, adb-3d-std, 
cool-2d-std, cool-2d-std-hden at $t=10^{15}$\,s.  For adb-3d-std the image is of a slice of the simulation 
at $z=0$.  It is clear that the magnetic field becomes highly distorted and amplified by the turbulence as 
it propagates through the precursor region. The units of field strength are $\mu$G, and the units of 
distance are $2.5\times10^{14}$\,cm (i.e.\ grid zone size).}
\end{center}
\end{figure*}

\begin{figure*}
\begin{center}
\includegraphics[width=15cm]{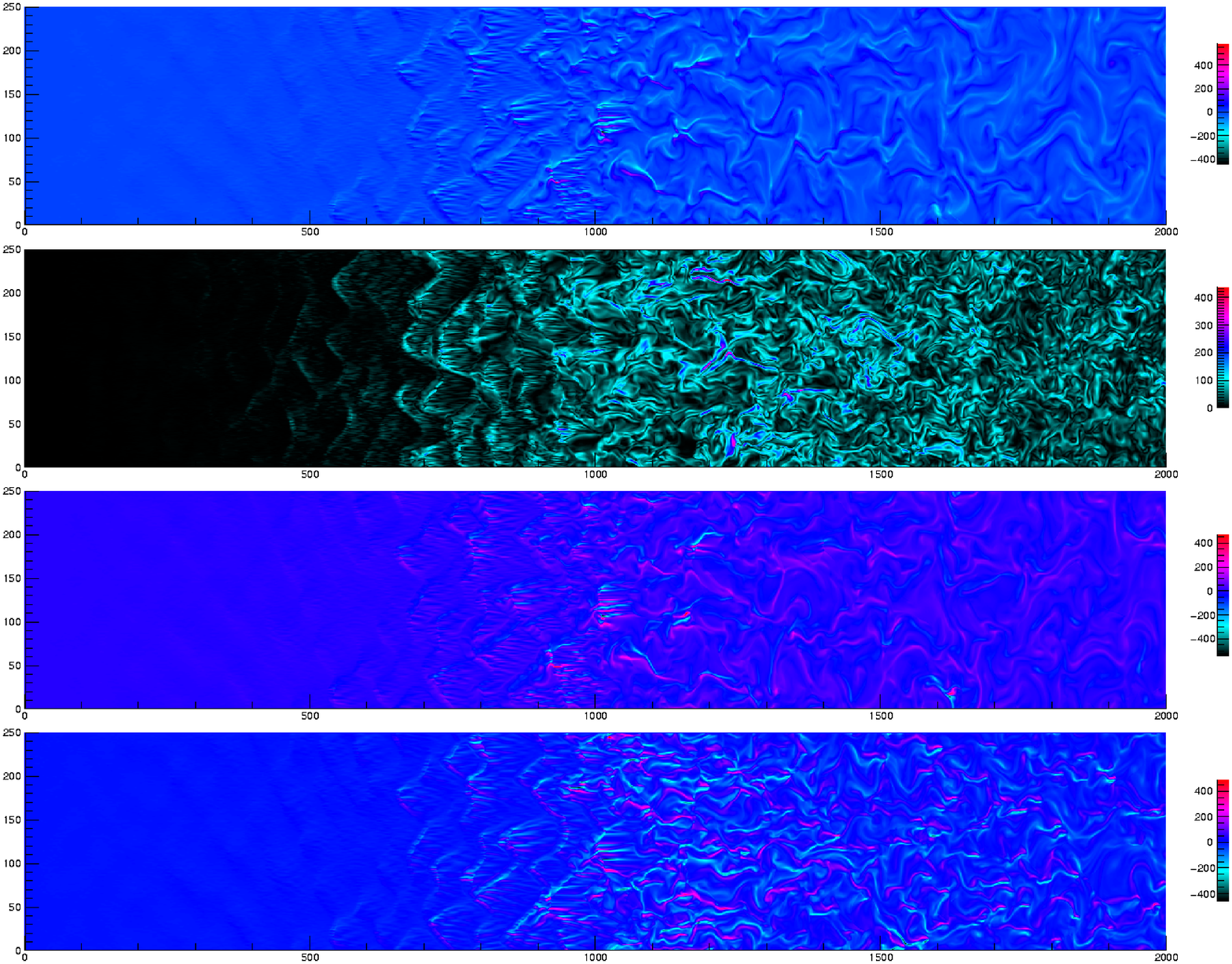}
\caption{ \label{fig:vort-image}  As Figure \ref{fig:bmag-image}, but for vorticity (normalised by
the flow time across the domain: $L/U_0$).  For adb-3d-std the magnitude of the vorticity is plotted
	for the slice of the simulation at $z=0$.  The vorticity tends to peak towards the centre of
the domain, then falls off as the ISM flows towards the right boundary.}
\end{center}
\end{figure*}

We now discuss the results of our simulations.  Figures \ref{fig:bmag-image} and \ref{fig:vort-image}
contain plots of the distribution of magnetic field strength and vorticity at $t=10^{15}$\,s for
simulations adb-2d-std, adb-3d-std, cool-2d-std and cool-2d-hden.  It is clear that the nature of the 
turbulence generated in each of the 2D simulations is rather similar, resulting in large-scale
structures while the turbulence in adb-3d-std involves somewhat smaller scale structures.  Even in
the latter case, the scales on which the field amplification occurs are still long in comparison to those 
of, for example, the Bell instability. The qualitative differences between 2D and 3D turbulence are 
discussed in more detail in Sect.\ \ref{sec:results-dim}.

Following the discussion in Sect.\ \ref{sec:introduction}, we split our results into discussions of the 
influence of dimensionality (Sect.\ \ref{sec:results-dim}), the orientation of the magnetic field (Sect.\ 
\ref{sec:results-angle}) and the influence of  radiative cooling (Sect.\ \ref{sec:results-cooling}).

\subsection{Influence of dimensionality}
\label{sec:results-dim}

We begin by examining the influence of dimensionality on the field 
amplification predicted by our toy model.  As mentioned above, this is 
an essential element in the study because of the qualitative differences
between 2D and 3D turbulence.  

\begin{figure}
\begin{center}
\includegraphics[width=9cm]{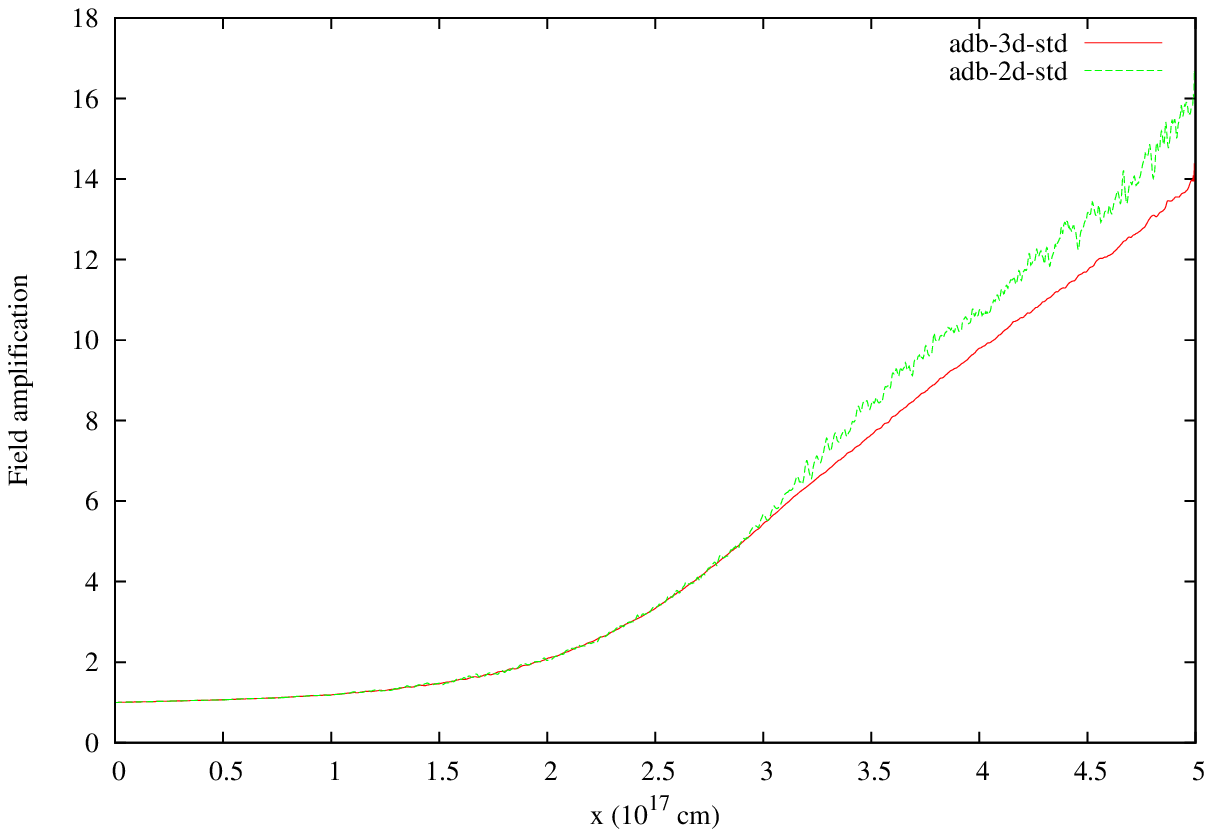}
\caption{ \label{fig:2d-3d-amp} Plots of the magnetic field
amplification, averaged over $0 \leq y \leq \frac{L}{8}$, for simulations
adb-2d-std and adb-3d-std. Note that the 3D simulation achieves slightly lower amplification than
the 2D one.}
\end{center}
\end{figure}

As can be seen from Fig.\ \ref{fig:2d-3d-amp} there is a difference in the amplification of the 
magnetic field achieved between the 2D and 3D simulations.  Significantly, the 2D system achieves 
a slightly higher (approximately 17\%) maximum amplification.  The extra variations seen in the plot
of amplification for adb-2d-std are due to the averaging being over a smaller number of 
grid zones than in adb-3d-std, since averaging in the latter is over a 2D area, while
averaging in the former is over a 1D line.

In order to understand the difference in the field amplification we can note that in a 3D 
(magneto)fluid system turbulent energy cascades from large scales to ever smaller scales until, 
finally, the viscous (or resistive) length scale is reached and the turbulent kinetic energy is then 
converted to heat. Hence the differential velocity, and therefore the turbulent kinetic energy, 
decreases with length scale.  Thus one expects that, unless the magnetic field is very weak, at some 
length scale the kinetic energy density associated with turbulent motion will fall below that of the 
magnetic energy density.  Below this length scale the fluid motions will not be able to efficiently 
stretch and twist the magnetic field, leading to a suppression of field amplification.  The same 
system in 2D, on the other hand, suffers a turbulent energy cascade from small to larger 
length scales (see Figure \ref{fig:bmag-image}).  Thus as the turbulence develops energy concentrates 
at large length scales, corresponding to high turbulent kinetic energy densities.  Hence, as the turbulence 
develops it is not suppressed as it is in 3D.  It is worth noting that this idealised picture of turbulence
should be considered only as an illustration: in fact the details of MHD turbulence are rather
different to that of hydrodynamic turbulence, and the turbulence in the system under
examination here is driven at {\em short} length scales.  However, the description above
suffices to give an intuitive understanding of the physical processes at work.

\begin{figure}
\begin{center}
\includegraphics[width=9cm]{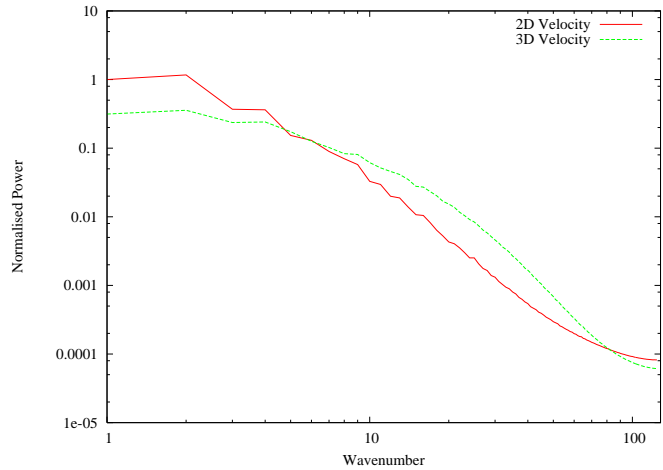}
\caption{ \label{fig:power-spec} Plots of the power spectrum of the velocity for the region of
the computational domain defined by $4.375\times10^{17}\mbox{\,cm} < x < 5 \times10^{17}$\,cm
for simulations adb-2d-std and adb-3d-std.}
\end{center}
\end{figure}

\begin{figure}
\begin{center}
\includegraphics[width=9cm]{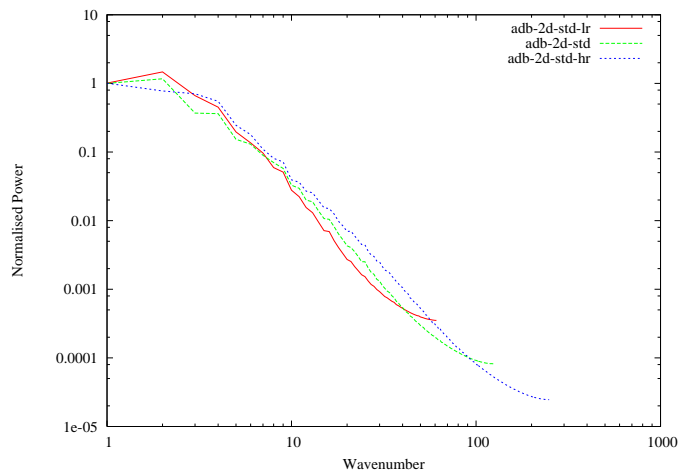}
\caption{ \label{fig:2d-power-res} Plots of the power spectrum of the velocity for the region of
the computational domain defined by $4.375\times10^{17}\mbox{\,cm} < x < 5 \times10^{17}$\,cm
for simulations adb-2d-std-lr and adb-2d-std and adb-2d-std-hr.}
\end{center}
\end{figure}

\begin{figure}
\begin{center}
\includegraphics[width=9cm]{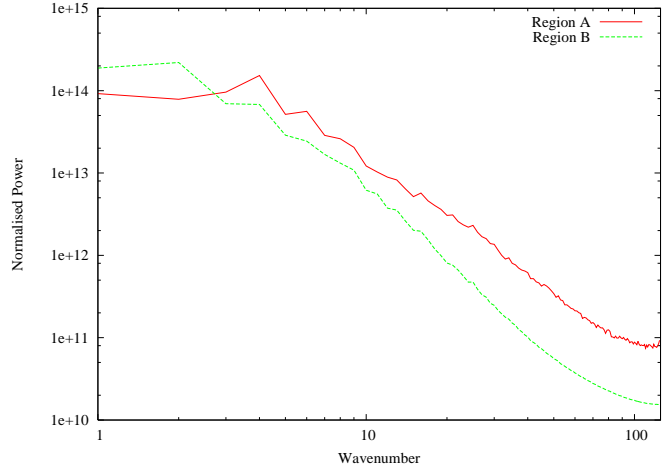}
\caption{\label{fig:power-spec-2d-3zone} Plots of the power spectrum of
the velocity for adb-2d-std in different regions of the simulations.
Region A is defined by $1.875\times10^{17}\,\mbox{cm} \leq x \leq 2.5 \times
10^{17}$\,cm and Region B is defined by
$4.375\times10^{17}\mbox{\,cm} < x < 5 \times10^{17}$\,cm.  An inverse
cascade is apparent, with the slope of the power spectrum in Region B
being steeper than that in Region A.}
\end{center}
\end{figure}

\begin{figure}
\begin{center}
\includegraphics[width=9cm]{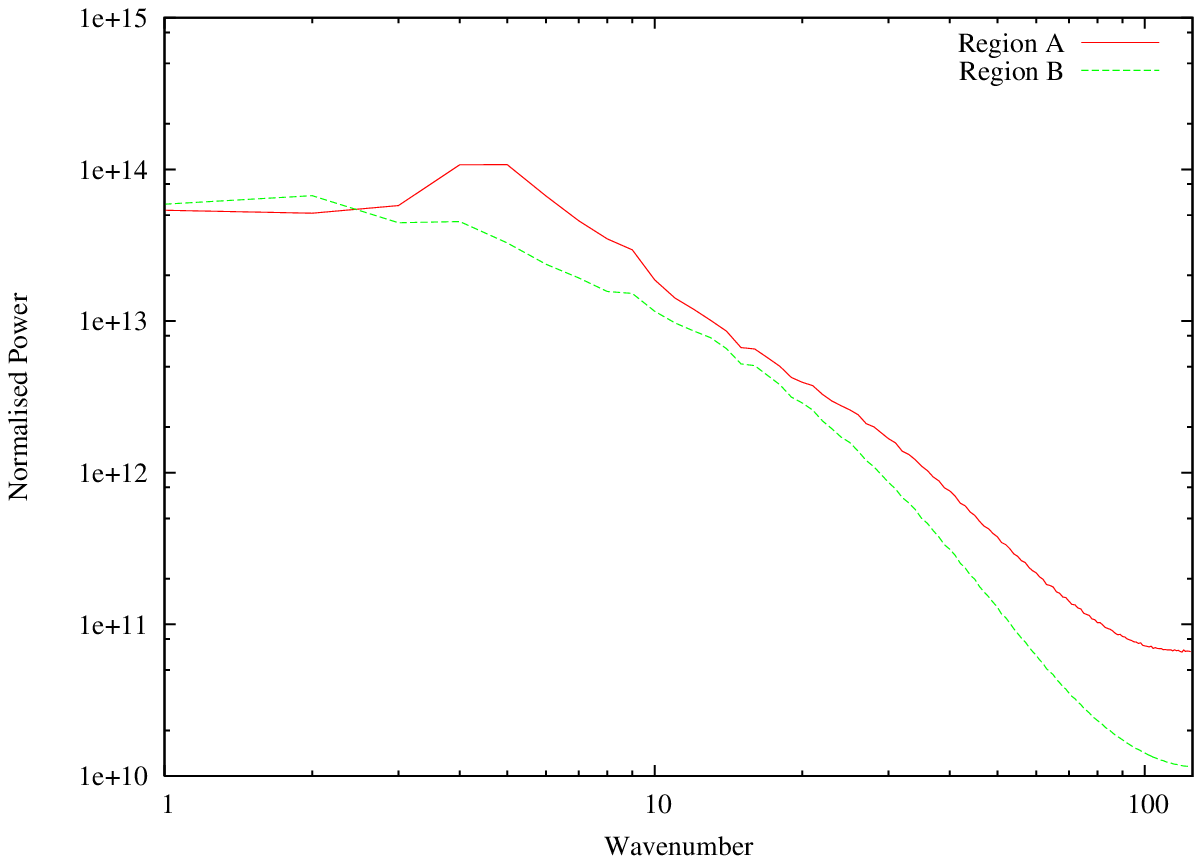}
\caption{\label{fig:power-spec-3d-3zone} As Fig.\
	\ref{fig:power-spec-2d-3zone} but for simulation adb-3d-std.  }
\end{center}
\end{figure}

To illustrate the differences (in $k$-space) between the 2D and 3D systems, Figure 
\ref{fig:power-spec} contains power spectra of the velocity in
simulations adb-2d-std and adb-3d-std in the final eighth of the grid (i.e.\ the region defined by 
$4.375\times10^{17}\mbox{\,cm} < x < 5 \times10^{17}$\,cm) where the turbulence will be well 
developed.  It is clear that at low wavenumbers (large length scales) there is more power in
the velocity variations in 2D than in 3D.  On the other hand, at intermediate wavenumbers
adb-3d-std has more power.  As noted in \citet{2012MNRAS.425.2277D}, numerical diffusion begins
to have a significant effect on velocity power spectra from the {\it HYDRA} code at wavelengths 
of less than about 10 -- 15 zones, corresponding to wavenumbers in Figure \ref{fig:power-spec} of 
greater than 15 -- 25.  Even so, it is clear that adb-3d-std exhibits more 
power at high $k$ than adb-2d-std, just as expected from general discussions 
of the differences between 2D and 3D turbulence.  To get some idea of the influence of
resolution on our power spectra, in Figure \ref{fig:2d-power-res} we present power 
spectra of the velocity in the region defined by $4.375\times10^{17}\mbox{\,cm} < x < 5 \times10^{17}$\,cm 
for adb-2d-std-lr, adb-2d-std and adb-2d-std-hr.  It is apparent that, at least in the range $k \leq
15$ the results from adb-2d-std and adb-2d-std-hr are similar, as expected.  Figures 
\ref{fig:power-spec-2d-3zone} and \ref{fig:power-spec-3d-3zone} illustrate the inverse cascade seen in 2D 
as opposed to 3D: the slope of the power spectrum in Region B ($4.375\times10^{17}\mbox{\,cm} < x < 5
\times10^{17}$\,cm) is steeper around $k=10$ than it is in Region A ($1.875\times10^{17}\,\mbox{cm} \leq 
x \leq 2.5 \times 10^{17}$\,cm) for adb-2d-std, while the reverse is true for simulation adb-3d-std.  
Hence, as the turbulence develops in the precursor region energy is transported across lengthscales in very 
much the way one would expect from a naive extension of incompressible hydrodynamic turbulence.

In summary, though, the dimensionality of the simulations, while clearly influencing the nature of the 
turbulence, only has a minor impact on the degree of magnetic field amplification 
achieved.  This is largely in agreement with the conclusion of \citet{2013MNRAS.436..294B} who 
detected no difference in amplification factors between 2D and 3D.  As described above,
however, one would expect a qualitative difference in the morphology of the amplified field,
and a small quantitative difference in the degree of amplification achieved, between 2D and 3D.

\subsection{Angle of magnetic field}
\label{sec:results-angle}

We now turn our attention to the influence of the direction of the magnetic
field amplification.  In \citetalias{2012MNRAS.427.2308D}, and for our standard simulations presented
here, we start with a magnetic field which is oriented perpendicular to the 
shock normal - i.e.\ we model a perpendicular shock.  Clearly this is the 
most favourable configuration for magnetic field amplification: the
differential acceleration resulting from the interaction of the ISM with
the cosmic ray pressure gradient immediately begins to shear the magnetic
field lines, leading to amplification.  If we consider the other extreme,
where we model a parallel shock, the differential acceleration (at least
initially) has no impact at all on the magnetic field strength.  

\begin{figure}
\begin{center}
\includegraphics[width=9cm]{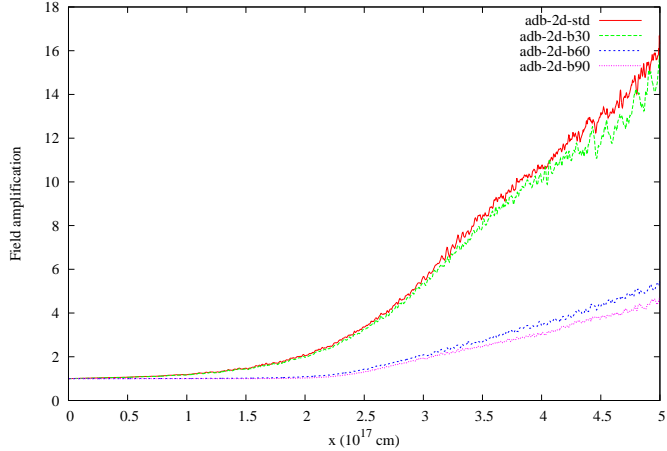}
\caption{ \label{fig:b-angle} Plots of the magnetic field
amplification, averaged over $0 \leq y \leq \frac{L}{8}$, for simulations
with different initial angles of the magnetic field to the direction of
flow.}
\end{center}
\end{figure}

Figure \ref{fig:b-angle} shows plots of the magnetic field amplification as
a function of distance for simulations adb-2d-b00, adb-2d-b30, adb-2d-b60
and adb-2d-std.  It is very clear that the angle between the initial (i.e.\
mean) magnetic field and the direction of flow has a large impact on the
field amplification achieved.  The results here are not unexpected.  To see
this let us make the somewhat radical simplification that only the transverse
component of the initial magnetic field contributes to the amplification.  
Then
\begin{equation}
\mathbf{B}_0 = B_0 \cos \alpha \, \mathbf{\hat{\imath}} + B_0 \sin \alpha \,
\mathbf{\hat{\jmath}}
\end{equation}
\noindent and
\begin{equation}
\mathbf{B}_{\rm f} = B_0 \cos \alpha \, \mathbf{\hat{\imath}} + F B_0 \sin \alpha \, 
\mathbf{\hat{\jmath}}
\end{equation}
\noindent where $\mathbf{B}_{\rm f}$ is the final magnetic field, $\alpha$ is the angle of the
field to the normal of the shock and $F$ can be measured from adb-2d-std (in which $\alpha=
\frac{\pi}{2}$).  We thus find that $F \approx 15$ and, defining $B_0 \equiv |\mathbf{B}_0|$
and $B_{\rm f} \equiv | \mathbf{B}_{\rm f}|$, so 
\begin{eqnarray}
\left.\frac{B_{\rm f}}{B_0}\right|_{\alpha = 60^\circ} \approx \left[0.75 + 
0.25F^2\right]^{1/2} = 7.54 \\
\left.\frac{B_{\rm f}}{B_0}\right|_{\alpha = 30^\circ} \approx \left[0.25 + 
0.75F^2\right]^{1/2} = 13
\end{eqnarray}
\noindent which is indeed roughly what is seen in Fig.\ \ref{fig:b-angle}.

Thus our radical simplification does indeed appear to reflect something of
the reality of the system: the transverse component of the magnetic field
dominates the effective amplification achieved.  Of course, there are 
clearly inaccuracies in this approximation: if there were none then simulation
adb-2d-b00 would exhibit no field amplification whatsoever.  The amplification
in this case arises from the fact that once the differential acceleration
has acted on the inflowing ISM bowshocks form around the denser components as
they plough through the less dense regions.  These bowshocks create a
transverse velocity component which, due to flux freezing, then also
creates a transverse magnetic field component.  This transverse component
can then be sheared in the same manner as the initial field in simulation
adb-2d-std, thereby becoming amplified.

One might reasonably wish to know the mean amplification achieved if the magnetic field in the
undisturbed medium is randomly oriented.  We can calculate this mean amplification factor, 
$\left<\frac{B_{\rm f}}{B_0}\right>$, as 
\begin{equation}
\left<\frac{B_{\rm f}}{B_0}\right> = \frac{2}{\pi} \int_0^{\frac{\pi}{2}} (\cos^2 \alpha +
F^2 \sin^2 \alpha)^{\frac{1}{2}}\,d\alpha \approx 9.6
\end{equation}
\noindent presuming the angle between the magnetic field and the shock normal is a uniformly 
distributed random variable.  This is likely to be approximately valid either for a supernova
encountering an ISM with a highly turbulent magnetic field, or for the mean amplification
measured over a large number of supernovae.

Thus, even allowing for variation in the orientation of the magnetic field, we can expect
amplification of at least a factor of 10 over what would be expected from the pre-existing ISM
magnetic field and standard shock physics.

\subsection{Influence of radiative cooling}
\label{sec:results-cooling}

One of the implications of our model is that an ensemble of shocks which are 
significantly weaker than the blastwave of the supernova remnant itself will
be formed as a result of the differential acceleration in the precursor.  
These shocks may be in the regime where significant radiative cooling can
occur via line emission.  This, at least in principle, raises the 
possibility that cooling itself may change the dynamics so as to enhance the magnetic field 
amplification.  This might naively be expected since cooling typically leads to greater compression
ratios (resulting from the radiating away of internal energy) and hence greater density contrasts, 
leading to greater differential acceleration and more rapid magnetic field amplification.  Since this 
is a rather complex system it is not clear, however, whether radiative cooling will actually 
have any effect at all.  We now turn our attention to studying the effect of including this in our 
simulations.

\begin{figure}
\begin{center}
\includegraphics[width=9cm]{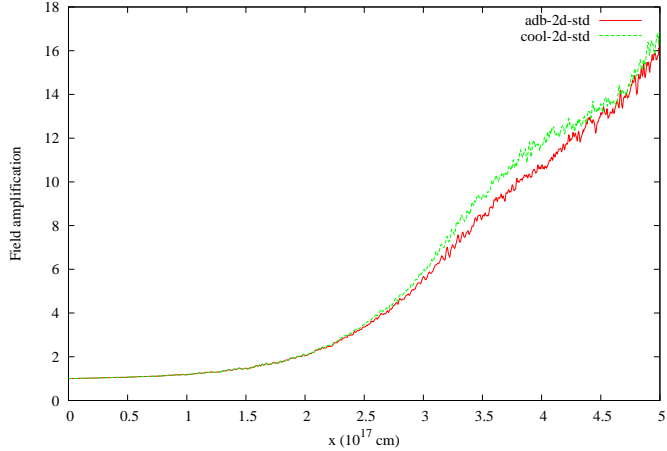}
\caption{ \label{fig:cool-std} Plots of the magnetic field
amplification for adb-2d-std and cool-2d-std.  It can be concluded that
the cooling makes rather little difference to the field amplification.}
\end{center}
\end{figure}

\begin{figure}
\begin{center}
\includegraphics[width=9cm]{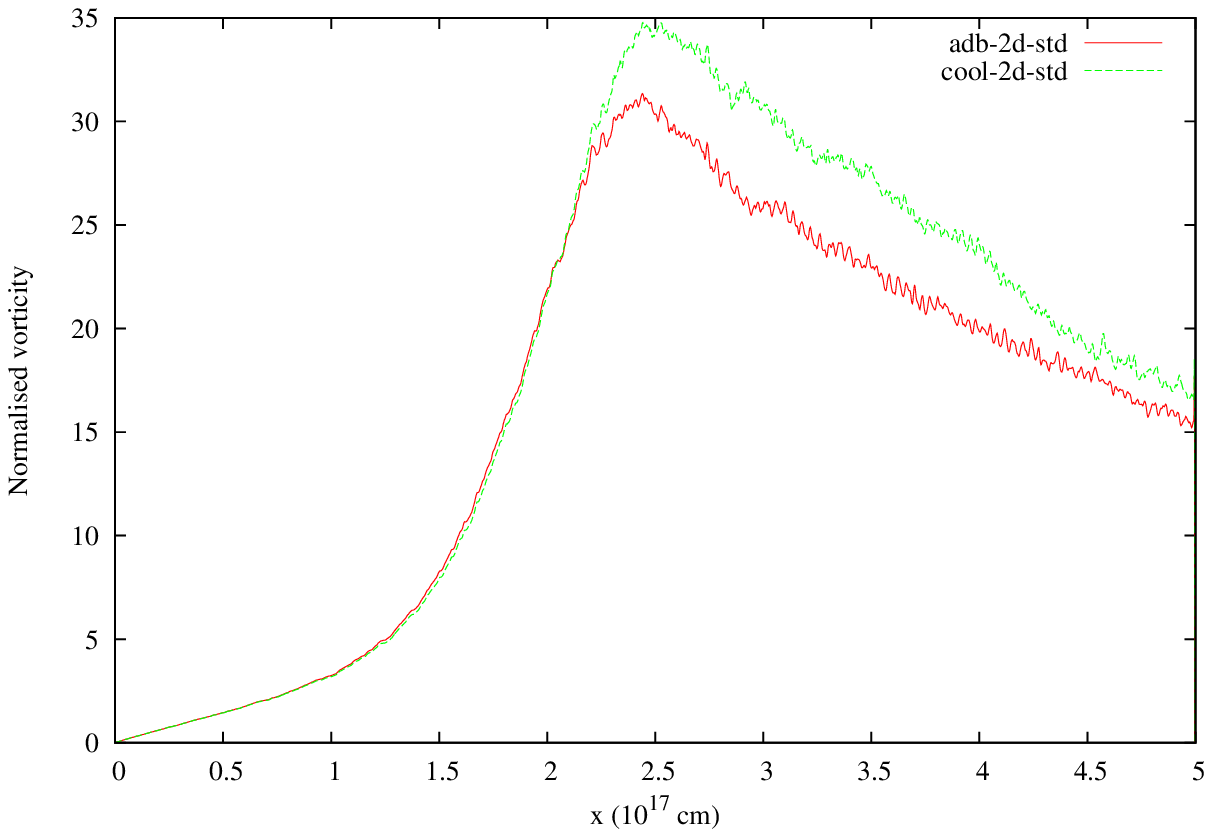}
\caption{ \label{fig:vort-cool-std} Plots of the vorticity normalised by the 
flow time through the precursor without deceleration (i.e.\
$\frac{L}{U_0}(\mathbf{\nabla}\times\mathbf{u}))$ for simulations adb-2d-std and cool-2d-std.}
\end{center}
\end{figure}

Figure \ref{fig:cool-std} contains plots of the magnetic field amplification
for adb-2d-std and cool-2d-std.  It is interesting to see that, in fact, the
addition of cooling via line emission has little impact on the overall
amplification factor.  A closer examination of Fig.\ \ref{fig:cool-std} gives
us some clue as to why this might be.  In the range $3\times10^{17} {\rm cm} \leq x \leq 4.5 \times 10^{17} {\rm cm}$ there is slightly enhanced 
amplification.  This enhancement disappears as we move towards $x = 
5\times10^{17}$\,cm.  This suggests that the cooling is most effective in
this part of the flow.  A further hint that this is the case is provided when
we examine the vorticity.  Figure \ref{fig:vort-cool-std} contains plots of
the vorticity (normalised by the flow-time through the precursor without 
deceleration) for simulations adb-2d-std and cool-2d-std.

\begin{figure}
\begin{center}
\includegraphics[width=9cm]{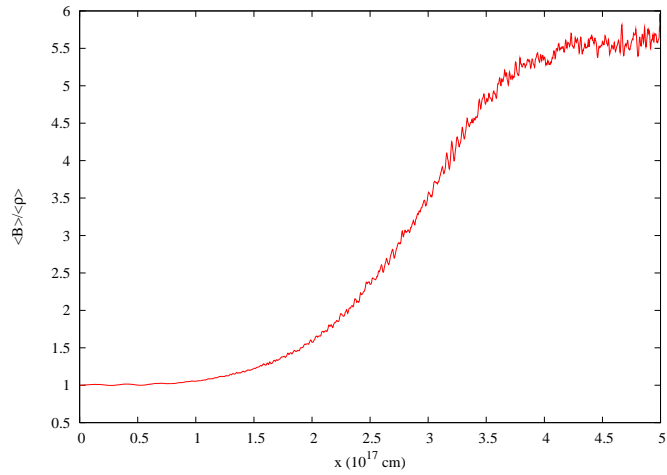}
\caption{ \label{fig:meanb-meanrho} Plot of $\frac{B}{\rho}$, averaged in the $y$ direction and
	normalised to its value at $x=0$, for adb-2d-std.  If compression were the dominant mechanism for
amplifying the field then this ratio would be approximately 1 for all $x$.}
\end{center}
\end{figure}

The vorticity in the flow contributes significantly to the
magnetic field amplification.  To see this, consider Figure \ref{fig:meanb-meanrho} which contains a 
plot of the ratio of the magnetic field strength to the mass density as a function of $x$.  If compression
were the dominant mode of strengthening the field this ratio should remain constant for all $x$.
In fact it increases to a value of 6, indicating that the magnetic field becomes 6 times stronger
than would be expected from compression alone.  As the ISM flows through the precursor the
vorticity grows and, in turn, this amplifies the field leading to the maximum
{\em growth} of the field occurring where there is high vorticity (cf Figs
\ref{fig:cool-std} and \ref{fig:vort-cool-std}).  It is clear from Fig.\ 
\ref{fig:vort-cool-std} that the vorticity does not grow monotononically as the 
flow proceeds through the precursor.  In fact there is a peak located
at approximately the centre of the computational domain, and the value 
of the vorticity at this peak is quite different in the cooled and adiabatic
simulations.  This difference in the vorticity is the cause of the enhanced
field amplification in cool-2d-std.  The vorticity decays as the flow
moves further through the precursor.  The vorticity causes efficient mixing
of the high and low density parts of the ISM, reducing differential
acceleration and leading to a reduction in the vorticity itself.  For these
parameters, where the initial magnetic field is weak in comparison to
the energy available from differential acceleration, this is the saturation 
mechanism of the instability.  Note that, physically, the lengthscales on which this mixing
occurs is defined by the viscous or resistive lengthscales.  In
numerical simulations, on the other hand, it is defined by numerical viscosity which, for the 
{\it HYDRA} code becomes noticeable below lengthscales of around 10 -- 15 zones (see Sect.\
\ref{sec:results-dim}). As noted in \citetalias{2012MNRAS.427.2308D}, mixing on unphysically large
lengthscales will {\em reduce} the field amplification. Therefore we expect
that the amplification achieved in our simulations is actually a {\em lower limit} for
the amplification which would be achieved in the precursor of a supernova.  We expect, however,
that the saturation mechanism of the instability will remain qualitatively the same.

\begin{figure}
\begin{center}
\includegraphics[width=9cm]{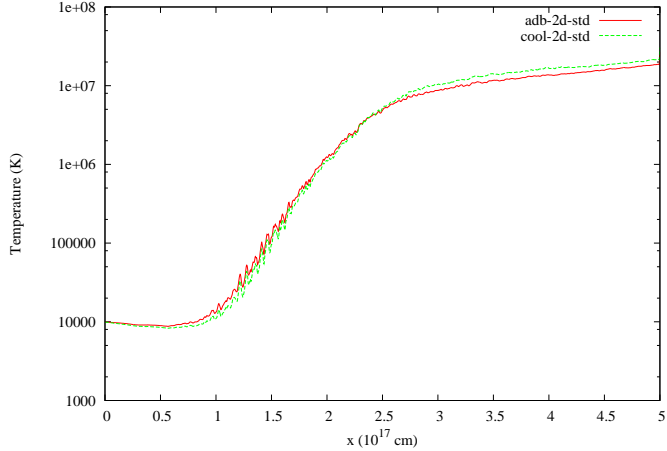}
\caption{ \label{fig:temp-cool-std} Plots of the mean temperature as a 
function of distance into the precursor for simulations adb-2d-std and 
cool-2d-std.}
\end{center}
\end{figure}

\begin{figure}
\begin{center}
\includegraphics[width=9cm]{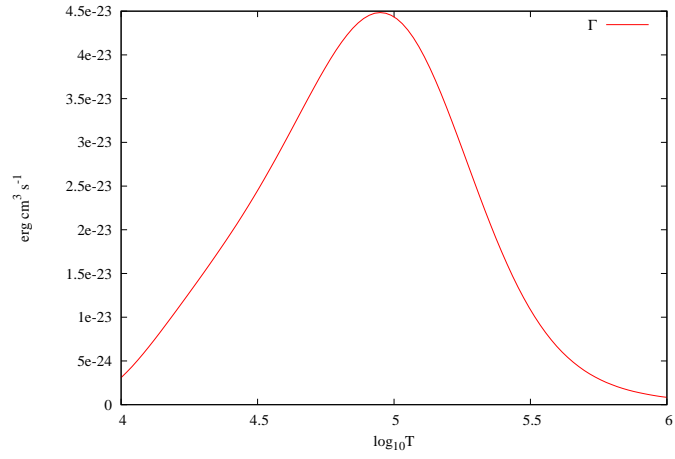}
\caption{ \label{fig:cooling-function} Plot of the cooling coefficient as a function of
temperature used in this work.}
\end{center}
\end{figure}

This mixing also reduces the influence of radiative cooling on the dynamics
of the process: the high density regions are mixed with low density
regions, and the shock strengths are reduced.  This is not the only reason for 
the lack of influence of line emission on the dynamics however.  Figure 
\ref{fig:temp-cool-std} shows plots of the mean temperature as a function of 
distance for adb-2d-std and cool-2d-std.  Figure \ref{fig:cooling-function} 
contains a plot of the cooling function used in this work.

A comparison of these two figures demonstrates clearly that the region
in which the fluid has a temperature at which cooling is significant 
(for example, the cooling coefficient is within an order of magnitude of the 
maximum) is between $1\times10^{17}$\,cm and $2\times10^{17}$\,cm.  To get an
idea of the likely influence of the cooling on the dynamics we calculate the
cooling time, the ratio of the internal energy density of the gas to the 
rate of loss of energy density due to radiative cooling:
\begin{equation}
\tau_{\rm c} = \frac{P}{(\gamma -1)n^2\Lambda(T)}
\end{equation}
Typical values in these simulations, with $n = 100$\pcm are between $10^{8}$\,s and $10^{9}$\,s.
In this time the gas will be advected a distance of around $\geq10^{17}$\,cm - a lengthscale
of order, or larger than, the time spent in the region on of the precursor in which the temperature
leads to appreciable cooling.

\begin{figure}
\begin{center}
\includegraphics[width=9cm]{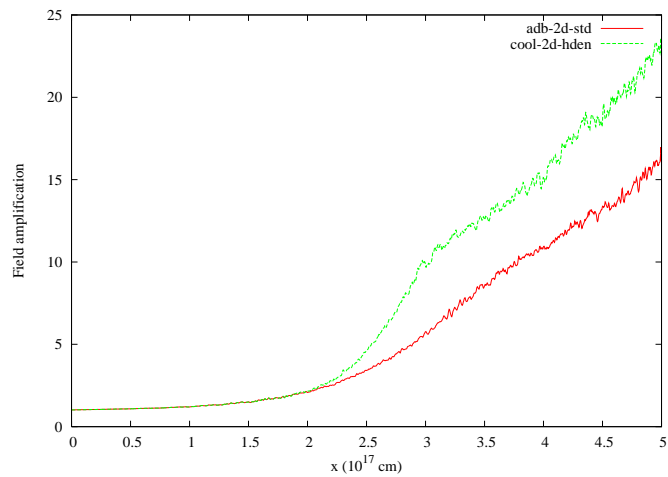}
\caption{ \label{fig:field-amp-hden} Plots of the field amplification as a function of distance for
	simulations adb-2d-std and cool-2d-hden.}
\end{center}
\end{figure}

To investigate this further, for higher cooling rates, we consider simulation adb-2d-hden.  Figure 
	\ref{fig:field-amp-hden} contains plots of the field amplification for adb-2d-std and
cool-2d-hden, the latter being a simulation in which the incoming ISM density is arbitrarily raised to a 
value of $10^3$\pcm.  Since the cooling we are describing here is collisional, this increases the
strength of the cooling by a factor of 100, and thus reduces the cooling length by the same factor.
Comparing with Figure \ref{fig:cool-std} we can see that, as expected, the cooling now leads to an 
enhanced amplification of the field.  The final amplification factor (at $x=5\times10^{17}$\,cm) is
approximately 23, instead of 16 for adb-2d-std.  Investigation of the mean compression of the flow in
these two simulations shows that each system results in mean compressions which are broadly similar,
indicating that the enhanced amplification comes about as a result of the increased vorticity
generated by the increased density contrasts in the radiatively cooled flow.  This suggests that 
supernova blastwaves encountering density enhancements in the ISM, such as molecular clouds, will exhibit 
enhanced magnetic field amplification.  Details of this mechanism are much more complicated in molecular 
cloud material, however, where cooling processes will include complex chemical networks and, possibly, 
multifluid effects.  This is the subject of future work.

\section{Conclusions}
\label{sec:conclusions}

We have investigated in some detail the model originally presented in 
\citetalias{2012MNRAS.427.2308D}, specifically with regard to the influence of the 
dimensionality of the turbulence, the angle of the magnetic field to the shock normal and the
likely influence of collisional cooling. The following summarises our main conclusions:
\begin{enumerate}
\item 3D turbulence leads to field amplification which is slightly ($\sim 17$\%) less than that 
achieved for 2D turbulence. We attribute this to the inverse cascade in 2D leading to a concentration 
of turbulent kinetic energy at large scales,
\item the angle of the magnetic field plays a critical role in the level of field amplification
achieved with values ranging between 5 and 17, depending on whether the shock normal is parallel 
or perpendicular to the prevailing magnetic field in the undisturbed ISM, 
\item mean values of amplification that are achieved can be expected to be at least 10, when
averaged over the possible angles between the magnetic field and the shock normal,
\item radiative cooling does not lead to appreciable impacts on the global dynamics of the
system, primarily due to the speed with which the ISM propagates through the precursor region
leading to the time it takes for the ISM to cool significantly being of order than, or longer than
its residence time in the precursor. This result is modified if the blastwave encounters density
enhancements, such as molecular clouds, in the ISM.  In this case the field amplification is
expected to be greater, though details of this mechanism in such an environment are complicated by
energetically significant chemical reaction networks and, possibly, multifluid effects.
\end{enumerate}

The model presented in \citetalias{2012MNRAS.427.2308D} appears to be robust to considerations such 
as the magnetic field orientation, the presence of collisional cooling and the dimensionality of the 
simulations.  Note that, in keeping with the philosophy of \citetalias{2012MNRAS.427.2308D}, we have 
neglected the details of the behaviour of the diffusion coefficient (e.g.\ its dependence on the field 
orientation).  Nonetheless, it seems that our overall conclusion, namely that this model can produce the
field amplification required to explain the observations, is generally applicable to SNRs in a 
variety of environments and with a variety of parameters.

\section*{Acknowledgements}
The authors wish to acknowledge the SFI/HEA Irish Centre for High-End 
Computing (ICHEC) for the provision of computational facilities and support.  TPD would like to
thank Aoife Curran for discussions of the power spectra of the simulations.  We would also like to
acknowledge the helpful comments of the anonymous referee.

\bibliographystyle{mn2e}

\bibliography{ArXiVversion-paper2}

\begin{thebibliography}{}

\bibitem[\protect\citeauthoryear{{Ballet}}{{Ballet}}{2006}]{2006AdSpR..37.1902%
B}
{Ballet} J.,  2006, Advances in Space Research, 37, 1902

\bibitem[\protect\citeauthoryear{{Bamba}, {Ueno}, {Nakajima} \&
  {Koyama}}{{Bamba} et~al.}{2004}]{2004ApJ...602..257B}
{Bamba} A.,  {Ueno} M.,  {Nakajima} H.,    {Koyama} K.,  2004, \apj, 602, 257

\bibitem[\protect\citeauthoryear{{Bamba}, {Yamazaki}, {Yoshida}, {Terasawa} \&
  {Koyama}}{{Bamba} et~al.}{2005}]{2005ApJ...621..793B}
{Bamba} A.,  {Yamazaki} R.,  {Yoshida} T.,  {Terasawa} T.,    {Koyama} K.,
  2005, \apj, 621, 793

\bibitem[\protect\citeauthoryear{{Begelman} \& {Zweibel}}{{Begelman} \&
  {Zweibel}}{1994}]{Begelman:1994fj}
{Begelman} M.~C.,  {Zweibel} E.~G.,  1994, \apj, 431, 689

\bibitem[\protect\citeauthoryear{{Bell}}{{Bell}}{2004}]{2004MNRAS.353..550B}
{Bell} A.~R.,  2004, \mnras, 353, 550

\bibitem[\protect\citeauthoryear{{Beresnyak}, {Jones} \&
  {Lazarian}}{{Beresnyak} et~al.}{2009}]{2009ApJ...707.1541B}
{Beresnyak} A.,  {Jones} T.~W.,    {Lazarian} A.,  2009, \apj, 707, 1541

\bibitem[\protect\citeauthoryear{{Berezhko}, {Ksenofontov} \&
  {V{\"o}lk}}{{Berezhko} et~al.}{2003}]{2003A&A...412L..11B}
{Berezhko} E.~G.,  {Ksenofontov} L.~T.,    {V{\"o}lk} H.~J.,  2003, \aap, 412,
  L11

\bibitem[\protect\citeauthoryear{{Brio} \& {Wu}}{{Brio} \&
  {Wu}}{1988}]{1988JCoPh..75..400B}
{Brio} M.,  {Wu} C.~C.,  1988, Journal of Computational Physics, 75, 400

\bibitem[\protect\citeauthoryear{{Br{\"u}ggen}}{{Br{\"u}ggen}}{2013}]{2013MNRA%
S.436..294B}
{Br{\"u}ggen} M.,  2013, \mnras, 436, 294

\bibitem[\protect\citeauthoryear{{Dahlburg} \& {Picone}}{{Dahlburg} \&
  {Picone}}{1989}]{1989PhFlB...1.2153D}
{Dahlburg} R.~B.,  {Picone} J.~M.,  1989, Physics of Fluids B, 1, 2153

\bibitem[\protect\citeauthoryear{{Dai} \& {Woodward}}{{Dai} \&
  {Woodward}}{1998}]{1998ApJ...494..317D}
{Dai} W.,  {Woodward} P.~R.,  1998, \apj, 494, 317

\bibitem[\protect\citeauthoryear{{Dedner}, {Kemm}, {Kr{\"o}ner}, {Munz},
  {Schnitzer} \& {Wesenberg}}{{Dedner} et~al.}{2002}]{Dedner:2002fr}
{Dedner} A.,  {Kemm} F.,  {Kr{\"o}ner} D.,  {Munz} C.-D.,  {Schnitzer} T.,
  {Wesenberg} M.,  2002, Journal of Computational Physics, 175, 645

\bibitem[\protect\citeauthoryear{{Downes}}{{Downes}}{2012}]{2012MNRAS.425.2277%
D}
{Downes} T.~P.,  2012, \mnras, 425, 2277

\bibitem[\protect\citeauthoryear{{Drury} \& {Downes}}{{Drury} \&
  {Downes}}{2012}]{2012MNRAS.427.2308D}
{Drury} L.~O.,  {Downes} T.~P.,  2012, \mnras, 427, 2308

\bibitem[\protect\citeauthoryear{{Drury} \& {Falle}}{{Drury} \&
  {Falle}}{1986}]{Drury:1986uq}
{Drury} L.~O.,  {Falle} S.~A.~E.~G.,  1986, \mnras, 223, 353

\bibitem[\protect\citeauthoryear{{Falle}, {Komissarov} \& {Joarder}}{{Falle}
  et~al.}{1998}]{1998MNRAS.297..265F}
{Falle} S.~A.~E.~G.,  {Komissarov} S.~S.,    {Joarder} P.,  1998, \mnras, 297,
  265

\bibitem[\protect\citeauthoryear{{Gaetz}, {Edgar} \& {Chevalier}}{{Gaetz}
  et~al.}{1988}]{1988ApJ...329..927G}
{Gaetz} T.~J.,  {Edgar} R.~J.,    {Chevalier} R.~A.,  1988, \apj, 329, 927

\bibitem[\protect\citeauthoryear{{Lagage} \& {Cesarsky}}{{Lagage} \&
  {Cesarsky}}{1983}]{1983A&A...125..249L}
{Lagage} P.~O.,  {Cesarsky} C.~J.,  1983, \aap, 125, 249

\bibitem[\protect\citeauthoryear{{Londrillo} \& {Del Zanna}}{{Londrillo} \&
  {Del Zanna}}{2000}]{2000ApJ...530..508L}
{Londrillo} P.,  {Del Zanna} L.,  2000, \apj, 530, 508

\bibitem[\protect\citeauthoryear{{Orszag} \& {Tang}}{{Orszag} \&
  {Tang}}{1979}]{1979JFM....90..129O}
{Orszag} S.~A.,  {Tang} C.-M.,  1979, Journal of Fluid Mechanics, 90, 129

\bibitem[\protect\citeauthoryear{{O'Sullivan} \& {Downes}}{{O'Sullivan} \&
  {Downes}}{2006}]{OSullivan:2006zr}
{O'Sullivan} S.,  {Downes} T.~P.,  2006, \mnras, 366, 1329

\bibitem[\protect\citeauthoryear{{O'Sullivan} \& {Downes}}{{O'Sullivan} \&
  {Downes}}{2007}]{OSullivan:2007mz}
{O'Sullivan} S.,  {Downes} T.~P.,  2007, \mnras, 376, 1648

\bibitem[\protect\citeauthoryear{{Parizot}, {Marcowith}, {Ballet} \&
  {Gallant}}{{Parizot} et~al.}{2006}]{2006A&A...453..387P}
{Parizot} E.,  {Marcowith} A.,  {Ballet} J.,    {Gallant} Y.~A.,  2006, \aap,
  453, 387

\bibitem[\protect\citeauthoryear{{Picone} \& {Dahlburg}}{{Picone} \&
  {Dahlburg}}{1991}]{1991PhFlB...3...29P}
{Picone} J.~M.,  {Dahlburg} R.~B.,  1991, Physics of Fluids B, 3, 29

\bibitem[\protect\citeauthoryear{{Ryu}, {Kang} \& {Jones}}{{Ryu}
  et~al.}{1993}]{Ryu:1993rt}
{Ryu} D.,  {Kang} H.,    {Jones} T.~W.,  1993, \apj, 405, 199

\bibitem[\protect\citeauthoryear{{Sutherland} \& {Dopita}}{{Sutherland} \&
  {Dopita}}{1993}]{1993ApJS...88..253S}
{Sutherland} R.~S.,  {Dopita} M.~A.,  1993, \apjs, 88, 253

\bibitem[\protect\citeauthoryear{{Uchiyama}, {Aharonian}, {Tanaka}, {Takahashi}
  \& {Maeda}}{{Uchiyama} et~al.}{2007}]{Uchiyama:2007ly}
{Uchiyama} Y.,  {Aharonian} F.~A.,  {Tanaka} T.,  {Takahashi} T.,    {Maeda}
  Y.,  2007, \nat, 449, 576

\bibitem[\protect\citeauthoryear{{Vink}}{{Vink}}{2012}]{2012A&ARv..20...49V}
{Vink} J.,  2012, \aapr, 20, 49

\bibitem[\protect\citeauthoryear{{Vink} \& {Laming}}{{Vink} \&
  {Laming}}{2003}]{2003ApJ...584..758V}
{Vink} J.,  {Laming} J.~M.,  2003, \apj, 584, 758

\bibitem[\protect\citeauthoryear{{V{\"o}lk}, {Berezhko} \&
  {Ksenofontov}}{{V{\"o}lk} et~al.}{2005}]{2005A&A...433..229V}
{V{\"o}lk} H.~J.,  {Berezhko} E.~G.,    {Ksenofontov} L.~T.,  2005, \aap, 433,
  229

\bibitem[\protect\citeauthoryear{{Webb}, {Zakharian} \& {Zank}}{{Webb}
  et~al.}{1999}]{Webb:1999vn}
{Webb} G.~M.,  {Zakharian} A.,    {Zank} G.~P.,  1999, Journal of Plasma
  Physics, 61, 553

\bibitem[\protect\citeauthoryear{{Yamazaki}, {Yoshida}, {Terasawa}, {Bamba} \&
  {Koyama}}{{Yamazaki} et~al.}{2004}]{2004A&A...416..595Y}
{Yamazaki} R.,  {Yoshida} T.,  {Terasawa} T.,  {Bamba} A.,    {Koyama} K.,
  2004, \aap, 416, 595

\end{thebibliography}

\begin{appendices}
\section{Sample Ideal MHD tests for \it{HYDRA}}
\label{sect:appendix}
The {\it HYDRA} code has been extensively tested for both ideal and multifluid MHD, as noted
in Sect.\ \ref{sec:method}.  The code is maintained using the {\tt subversion}\footnote{See {\tt http://subversion.apache.org/}.} versioning system and each
night the latest version of the code is automatically downloaded from the main repository and tested 
against a
number of tests.  Each week a full suite of tests is run on the latest version of the code.  In this
section we provide two of the ideal MHD tests we use to demonstrate the effectiveness of
{\it HYDRA} in simulating ideal MHD systems.  The Brio-Wu shocktube test demonstrates the
shock-capturing nature of the code, while the Orszag-Tang vortex is useful to test the
multidimensional performance of the code.  Both of these are standard tests used in the literature.

\subsection{Brio-Wu Shocktube test}
\label{sec:brio-wu}
This is a 1D test proposed by \citet{1988JCoPh..75..400B} in which a left and right state are defined and 
are initially separated by a discontinuity.  Once the system begins to evolve this discontinuity breaks down 
into a series of shocks and rarefactions of waves of different types.  Using the notation that 
$\mathbf{u}$ is the state vector, with $\mathbf{u} = (\rho, u_x, u_y, u_z, P, B_x, B_y, B_z)^{\rm T}$, 
the initial conditions are as follows:
\begin{equation}
\mathbf{u}_{\rm L} = \left(\begin{array}{c} 1 \\ 0 \\ 0 \\ 0 \\ 1 \\ 0.75 \\ 1 \\ 0 \end{array}
		\right) \hfill \mathbf{u}_{\rm R} = \left(\begin{array}{c} 0.125 \\ 0 \\ 0 \\ 0 \\ 0.1 \\ 0.75 \\ -1
			\\ 0 \end{array}
			      \right).
\end{equation}
The initial discontinuity is positioned at $x=0.5$ and we use 1000 grid points in the test presented
here.

\begin{figure}
\begin{center}
\includegraphics[width=9cm]{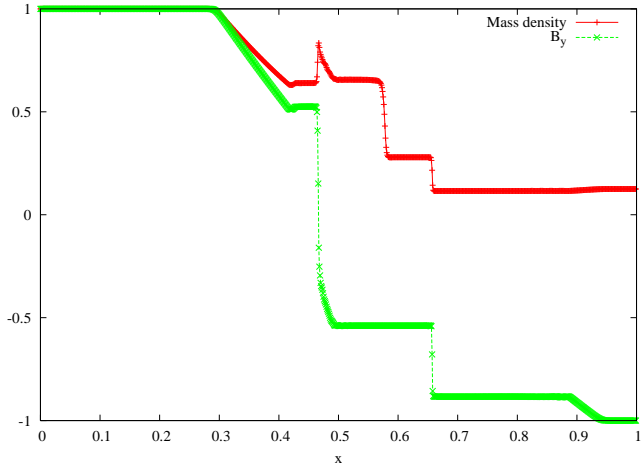}
\caption{ \label{fig:brio-wu} Plots of the mass density and the $y$ component of the magnetic
	field at $t=0.12$ for the Brio-Wu shocktube test.}
\end{center}
\end{figure}

Figure \ref{fig:brio-wu} contains plots of the density, $\rho$, and the $y$ component of the
magnetic field as a function of $x$ at time $0.12$.  These results can be compared with
those presented in, for example, \citet{1998MNRAS.297..265F}.  It can be seen that the results match those 
in the literature rather well.

\subsection{The Orszag-Tang Vortex test}

This test was first studied by \citet{1979JFM....90..129O} for the incompressible case, and
subsequently by \citet{1989PhFlB...1.2153D} and \cite{1991PhFlB...3...29P} for the compressible case
presented here. It is widely used as a test of the performance of multidimensional MHD codes in tracking 
the evolution of shocks and the transition to turbulence of an initially non-turbulent medium.  The 
initial conditions used are
\begin{equation}
\mathbf{u} = \left(\begin{array}{c}
		       \frac{25}{36 \pi} \\
				 \sin(2 \pi y) \\
				 \sin(2 \pi x) \\
				 0 \\
				 \frac{5}{12 \pi} \\
				 - \frac{1}{(4 \pi)^{1/2}} \sin (2 \pi y) \\ 
				 \frac{1}{(4 \pi)^{1/2}} \sin (4 \pi x) \\ 
				 0 \end{array} \right)
\end{equation}
\noindent using the notation defined in \ref{sec:brio-wu} for the definition of the state vector. The
simulation is run on a unit square using $512\times512$ grid zones and periodic boundary conditions in all 
directions.

\begin{figure}
\begin{center}
\includegraphics[width=9cm]{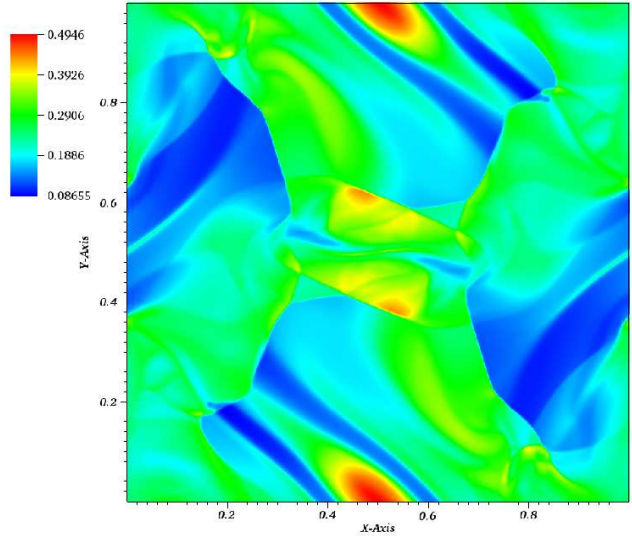}
\caption{ \label{fig:orszag-tang-image} Distribution of mass density in the Orszag-Tang vortex test at
$t=0.5$.}
\end{center}
\end{figure}

\begin{figure}
\begin{center}
\includegraphics[width=9cm]{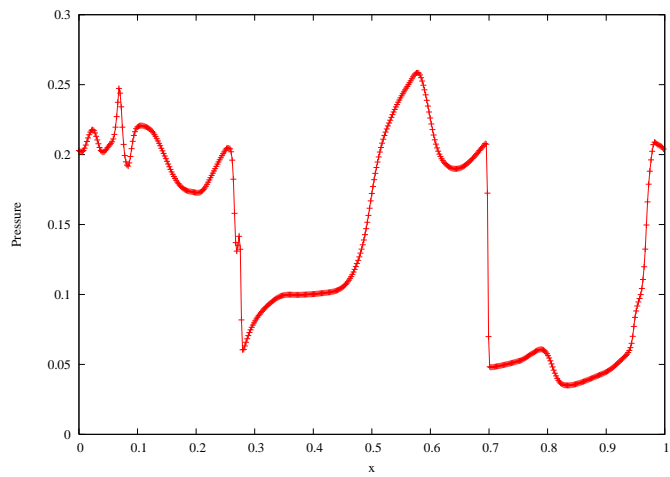}
\caption{ \label{fig:orszag-tang-plot} Plot of the thermal pressure in the Orszag-Tang vortex test at
	$t=0.5$ and $y=0.3125$.
temperature used in this work.}
\end{center}
\end{figure}

Figure \ref{fig:orszag-tang-image} contains a plot of the distribution of the mass density at $t=0.5$.
This can be seen to compare well with Figure 6 of \citet{1998ApJ...494..317D} and Figure 10 of
\citet{2000ApJ...530..508L}.  Figure \ref{fig:orszag-tang-plot} contains a plot of the pressure at 
$y=0.3125$ at $t=0.5$.  This plot can be compared with the lower panel in Figure 11 of 
\citet{2000ApJ...530..508L}.  In all cases the results of {\it HYDRA} for ideal MHD can be seen to 
match well with those published by other authors using a variety of codes and algorithms, indicating
that we can have some confidence in the results presented here.
\end{appendices}

\end{document}